\documentclass[11pt]{article}

\usepackage{cite}
\usepackage{amssymb}
\usepackage{amsmath}
\usepackage{bm}
\usepackage{enumerate}
\usepackage{graphicx}
\usepackage{epstopdf}
\begin{document}

\title{\bf Bohm's approach and individuality.}
\author{P. Pylkk\"{a}nen$^ {1, 2}$, B. J. Hiley$^{3}$ and I. P\"{a}ttiniemi$^1$.}
\date{{\small 1. Department of Philosophy, History, Culture and Art Studies \& The Finnish Center of Excellence in the Philosophy of the Social Sciences (TINT), University of Helsinki, Finland\\ 2. Department of Cognitive Neuroscience and Philosophy, University of Sk\"{o}vde, Sweden \\3.  TPRU, Birkbeck, University of London, Malet Street, \\London WC1E 7HX.}\\ \vspace{0.4cm}(\small 15 November 2014)}
\maketitle

\begin{abstract}
Ladyman and Ross (LR) argue that quantum objects are not individuals (or are at most weakly discernible individuals) and use this idea to ground their metaphysical view, ontic structural realism, according to which relational structures are primary to things.  LR acknowledge that there is a version of quantum theory, namely the Bohm theory (BT), according to which particles do have definite trajectories at all times. However, LR interpret the research by Brown  {\em et al.}  as implying that ``raw stuff" or {\em haecceities} are needed for the individuality of particles of BT, and LR dismiss this as idle metaphysics. In this paper we note that Brown {\em et al.}'s  research does not imply that {\em haecceities} are needed. Thus BT remains as a genuine option for those who seek to understand quantum particles as individuals.  However, we go on to discuss some problems with BT which led Bohm and Hiley to modify it.  This modified version underlines that, due to features such as context-dependence and non-locality, Bohmian particles have a very limited autonomy in situations where quantum effects are non-negligible.  So while BT restores the possibility of quantum individuals, it also underlines the primacy of the whole over the autonomy of the parts.  The later sections of the paper also examine the Bohm theory in the general mathematical context of symplectic geometry.  This provides yet another way of understanding the subtle, holistic and dynamic nature of Bohmian individuals. We finally briefly consider Bohm's other main line of research, the `"mplicate order", which is in some ways similar to LR's structural realism. \end{abstract}

\section{Introduction.}

\begin{quote}
The usual interpretation of the quantum theory implies that we must renounce the possibility of describing an individual system in terms of a single, precisely defined conceptual model. We have, however, proposed an alternative interpretation ...which leads us to regard a quantum-mechanical system as a synthesis of a precisely definable particle and a precisely definable $\psi$-field... (Bohm 1952a: 188)\end{quote}

Perhaps the greatest challenge to the notion that objects are individuals with well-defined identity conditions comes from modern quantum and relativity physics. For, ever since the early days of the quantum revolution, the identity and individuality of quantum systems has frequently been called into question (see e.g. French 2011 and the references therein; French and Krause 2006; Ladyman and Ross 2007, ch 3). 

Many of the founding figures of quantum theory, and most notably Niels Bohr, held that it is not possible to describe individual quantum objects and their behaviour in the same way as one can in classical physics, pointing out  that the individual cannot be separated from the whole experimental context (for a recent penetrating discussion of Bohr's views, see Plotnitsky 2010). The idea that quantal objects might, in some sense, be ``non-individuals" was also considered early on by, for example, Born, Heisenberg and Weyl (French 2011: 6).

One physicist who throughout his career emphasized the holistic features of quantum phenomena was David Bohm (1917-1992). For example, in his 1951 text-book  {\em Quantum Theory} which reflected the usual interpretation of quantum theory, he characterized individual quantum objects in strongly relational and contextual terms: 

\begin{quote}...quantum theory requires us to give up the idea that the electron, or any other object has, by itself, any intrinsic properties at all. Instead, each object should be regarded as something containing only incompletely defined potentialities that are developed when the object interacts with an appropriate system (1951: 139).
\end{quote}

However, as is well known, soon after completing his 1951 text-book, Bohm discovered an alternative interpretation of quantum theory which gives individuals a much stronger status than the usual interpretation. His motivation stemmed from his dissatisfaction with the fact that the usual interpretation was not providing an ontology, a comprehensive view of quantum reality beyond the fragmentary experimental phenomena (Bohm 1987).  Besides, discussions with Einstein in Princeton in the early 1950s strongly inspired him to start searching for a deterministic extension of quantum theory:

\begin{quote} In this connection, I soon thought of the classical Hamilton-Jacobi theory, which relates waves to particles in a fundamental way.  Indeed, it had long been known that when one makes a certain approximation [WKB], Schr\"{o}dinger's equation becomes equivalent to the classical Hamilton-Jacobi equation. At a certain point, I suddenly asked myself: What would happen, in the demonstration of this equivalence, if we did not make this approximation? (Bohm 1987)\end{quote}

From the ontological point of view the puzzling thing about the WKB approximation is that we start from Schr\"{o}dinger's equation, which according to the usual interpretation does not refer to an ontology, we then remove something from Schr\"{o}dinger's equation, and suddenly obtain the classical Hamilton-Jacobi equation which refers to a classical ontology.  But how can {\em removing} something from a ``non-ontology" give us an ontology? Bohm's insight was to realize that if one does not approximate, one can see an unambiguous ontology that is hiding in Schr\"{o}dinger's equation:

\begin{quote} 	 I quickly saw that there would be a new potential, representing a new kind of force, that would be acting on the particle. I called this the quantum potential, which was designated by {\em Q}. This gave rise immediately to what I called a causal interpretation of the quantum theory. The basic assumption was that the electron {\em is} a particle, acted on not only by the classical potential, {\em V}, but also by the quantum potential, {\em Q}.  This latter is determined by a new kind of wave that satisfies Schr\"{o}dinger's equation. This wave was assumed, like the particle, to be an independent actuality that existed on its own, rather than being merely a function from which the statistical properties of phenomena could be derived.
\end{quote}  

{\em Q} is responsible for all quantum effects (such as the interference patterns of electrons and quantum nonlocality).  However, whenever {\em Q} is negligibly small, quantum ontology gives rise to classical ontology. In 1952 Bohm published in {\em Physical Review} two papers that presented this interpretation (which independently re-discovered and made more coherent a theory which de Broglie had presented in the 1927 Solvay conference). To see the relevance of Bohm's interpretation to the question of individuality in quantum theory let us consider how he contrasts his approach with that of Bohr:
 
\begin{quote}...Bohr suggests that at the atomic level we must renounce our hitherto successful practice of conceiving of an individual system as a unified and precisely definable whole, all of whose aspects are, in a manner of speaking, simultaneously and unambiguously accessible to our conceptual gaze. ... in Bohr's point of view, the wave function is in no sense a conceptual model of an individual system, since it is not in a precise (one-to-one) correspondence with the behavior of this system, but only in a statistical correspondence (1952a: 167-8).
\end{quote}

In contrast to this, Bohm's alternative interpretation regards
 
 \begin{quote} ...the wave function of an individual electron as a mathematical representation of an objectively real field (1952a: 170).\end{quote}
 
Thus for Bohm, an individual quantum-mechanical system has two aspects:
  
  \begin{quote} it is a synthesis of a precisely definable particle and a precisely definable $\psi$-field which exerts a force on this particle (1952b: 188).  
  \end{quote}
  
Now, if the Bohm theory is a coherent option, it undermines the arguments of those who claim that non-relativistic quantum theory somehow forces us to give up the notion that quantum objects are individuals with well-defined identity conditions. Ironically there is also a tension between Bohm's 1952 theory and much of his own other more anti-individualist (i.e. structuralist and process-oriented) work - both before and after 1952.  Given this tension, it is not surprising that Bohm and Hiley developed Bohm's 1952 theory further in their later research (Bohm and Hiley 1987, 1993).

The question of whether quantum particles are individuals is also raised by the philosophers James Ladyman and Don Ross in their thought-provoking and important book  {\em Every Thing Must Go} (2007, hereafter ETMG). They advocate the view that quantum particles are not individuals (or, at most, are weakly discernible individuals). They acknowledge that there seem to be individuals in the Bohm theory, but go on to refer to research by Brown  {\em et al.}  (1996) which they interpret as saying that in the Bohm theory, the properties normally associated with particles (mass, charge, etc.) are inherent only in the quantum field and not in the particles.  It would then seem that there is nothing there in the trajectories unless one assumes the existence of some ``raw stuff" of the particle. In other words it seems that {\em haecceities} are needed for the individuality of particles of the Bohm theory, and Ladyman and Ross dismiss this as idle metaphysics.

In what follows we will first give a brief account of Ladyman and Ross's views on individuality in quantum mechanics (section 2).  We then introduce Bohm's 1952 theory, focusing on the way it seems to have room for quantum individuals (section 3).   In section 4 we first present Ladyman and Ross's criticism of Bohmian individuality and then go on to challenge this criticism. We illustrate how puzzling quantum experiments, such as the Aharonov-Bohm effect are explained in terms of the quantum potential approach.  In section 5 we show how the symplectic symmetry forms a basis common to both classical and quantum motion and in section 6 bring out its relevance to the question of quantum individuality.  In section 7 we show how some of the problems with the 1952 Bohm theory can be resolved by the radical proposal that the quantum potential functions as active information.  We also note that while the Bohm theory allows us to retain the notion of individual particles, such particles have only a limited autonomy.   In section 8 we briefly consider Bohm's other main line of research (the ``implicate order") which emphasizes the primacy of structure and process over individual objects. This research further underlines that while individuals have relative autonomy in Bohm's approach, they are not fundamental. Thus, in a broad sense, Bohm and Hiley's approach to quantum theory has interesting similarities to Ladyman and Ross's structural realism, even though the former gives quantum individuals a stronger status than the latter.
 
\section{Ladyman and Ross on individuality in quantum mechanics.}

In the third chapter of ETMG, Ladyman and Ross discuss identity and individuality in quantum mechanics. Following French and Redhead (1988), they first establish that indistinguishable elementary particles, that is particles that have the same mass, charge etc., behave differently in quantum mechanics than they do in classical statistical mechanics. For quantum particles an  ``indistinguishability postulate" states that a permutation of indistinguishable particles is not observable and thus those states which differ only by a permutation of such particles are treated as the same state with a different labeling. This might point to the view that quantum particles are not individuals.

Individuality is however an ontological property whereas (in)distinguishability is an epistemic one. So how are these two related? Ladyman and Ross identify three candidates in the philosophical tradition for individuality:
\begin{enumerate}
\item transcendent individuality: the individuality of something is a feature of it over and above all its qualitative properties;
\item spatio-temporal location or trajectory;
\item all or some restricted set of their properties (the bundle theory) (ETMG, p. 134).
\end{enumerate}

$\#1$ above is ruled out because it involves {\em haecceities}, and thus involves what Ladyman and Ross would consider idle metaphysical speculation. Granting this restriction for the sake of the argument, the interesting candidates are $\#2$ and $\#3$

A connection between individuality and distinguishability is given by the Principle of the Identity of Indiscernibles (PII), which can be taken roughly to state that no two objects have exactly the same properties. It is easy to see that everyday objects satisfy both $\#2$ and $\#3$, while the point particles of classical mechanics satisfy $\#2$. Then for both everyday objects and particles of classical mechanics PII is true and individuality and distinguishability can be taken to be the same thing. However, for certain quantum systems neither $\#2$ nor $\#3$ seems to hold. Ladyman and Ross take as an example of such a state the singlet state of two electrons orbiting a helium atom:
\begin{eqnarray}
\psi =1/\sqrt 2 [|\uparrow\rangle_1 |\downarrow\rangle_2 - |\downarrow\rangle_1 |\uparrow\rangle_2)] 		\label{eq:tang}
\end{eqnarray}

Here any property that can be ascribed to particle 1 can also be ascribed to particle 2. So in this state the two electrons share all their extrinsic and intrinsic properties, thus falling foul of both $\#2$ and $\#3$. So it would seem that quantum particles are not individuals.

However, this conclusion might follow from a too strict a notion of discernibility. Following Saunders (2003a, 2003b, 2006), Ladyman and Ross give three notions of discernibility: 
\begin{enumerate}[(i)]
\item
 absolute discernibility 
\item
relative discernibility and 
\item
 weak discernibility. 
\end{enumerate}

These can be defined as follows (ETMG, p 137):
\begin{enumerate}[(i)]
\item
``Two objects are absolutely discernible if there exists a formula in one variable which is true of one object and not the other". This holds for ordinary everyday objects.
\item
``Two objects are relatively discernible just in case there is a formula in two free variables which applies to them in one order only. ...[T]he points of a one-dimensional space with an ordering relation, since, for any such pair of points $x$ and $y$, if they are not the same point then either $x > y$ or $x < y$ but not both".
\item
``Two objects are weakly discernible just in case there is two-place irreflexive relation that they satisfy."  The Fermions in a singlet state are discernible in this sense, as they satisfy the relation `is of opposite spin to'.
\end{enumerate}

Now since electrons in the singlet state are discernible they can be viewed as individuals. But they are weakly discernible. This is a thoroughly structuralist view ``...as individuals are nothing over and above the nexus of relations in which they stand." (ETMG, p. 138.)

\section{ The Bohm theory}

Now let us turn to consider how the Bohm theory deals with these situations. Starting from the Schr\"{o}dinger equation, we find the real part can be written as
\begin{eqnarray}
\frac{\partial S}{\partial t}+\frac{1}{2m}(\nabla S)^2+Q+V=0	\label{eq:qhj}
\end{eqnarray}
under a polar decomposition of the wave function $\psi(\bm r,t)=R(\bm r,t)\exp[iS(\bm r,t)/\hbar]$.
This equation has a similar form to the classical Hamilton-Jacobi equation except for the appearance of a new term
\begin{eqnarray}
Q=-\frac{\hbar^2}{2m}\frac{\nabla^2R}{R}
\end{eqnarray}
which is known as the quantum potential.  Of course this means identifying the momentum of the particle by $p=\nabla S$ where $S$ is the phase of the wave function. This relation, also known in some versions of this approach as the `guidance condition', enables the trajectory of the particle to be calculated. Note that the so called ``Bohmian mechanics" approach to Bohm's theory emphasizes that we get a deterministic particle mechanics directly from the first-order guidance equation involving the velocities of the particles (see Goldstein 2013). However, in this paper we will be focusing on the way the Bohm theory arises from the above Hamilton-Jacobi type equation.

Figures 1 and 2 provide well-known visualizations.

\begin{figure}[h]
   \centering
    \includegraphics[width=2in]{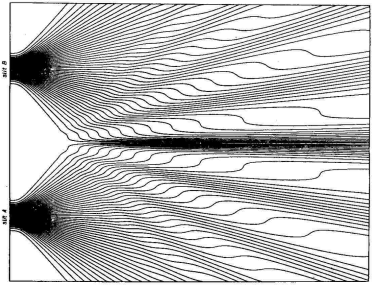}
   \caption{Trajectories for two Gaussian slits}
   \label{fig:2}
\end{figure}

\begin{figure}[h]
   \centering
    \includegraphics[width=2in]{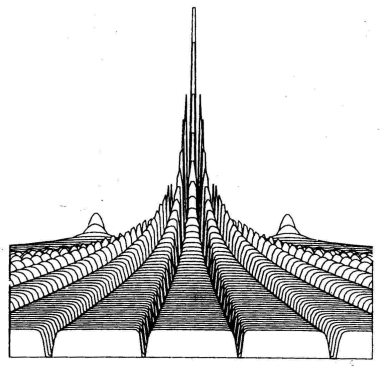} 
   \caption{Quantum potential for two Gaussian slits}
   \label{fig:1}
\end{figure}

So there is a version of quantum theory (the Bohm theory) according to which each particle has a definite and distinct trajectory at all times. This suggests that quantum particles are individuals, with position being the property in virtue of which particles are always different from one another. 

The biggest problem for retaining the notion of individuality is particles in entangled states described by equation (\ref{eq:tang}).  This is an entangled spin state which has been discussed in detail by Dewdney, Holland, Kyprianidis and Vigier (1988), but for our purposes here it is sufficient to consider a general two-body wave function, $\psi(\bm r_1, \bm r_2, t)$ 
and use the two-body Schr\"{o}dinger equation to find
\begin{eqnarray}
\frac{\partial S}{\partial t}+\frac{(\nabla_1S)^2}{2m_1}+\frac{(\nabla_2S)^2}{2m_2}+Q(\bm r_1, \bm r_2,t)+V(\bm r_1, \bm r_2)=0
\end{eqnarray}
The second and third terms in this equation correspond to the kinetic energies of each particle so once again the problem of individuality does not seem to arise in the Bohm theory.   The entanglement is reflected in the non-local quantum potential energy term $Q(\bm r_1, \bm r_2,t)$.  Furthermore since the trajectories do not cross, we follow Brown, Sj\"{o}qvist and Bacciagaluppi (1999: 233) and conclude that indistinguishable fermions will  always have distinct trajectories (for further discussion see French 2011: 14-15; French and Krause 2006: 178).  Thus individuality is preserved.

\section{The Bohm theory and {\em haecceities}}

So the Bohm theory seems to suggest, {\em contra} Ladyman and Ross, that quantum particles can be individuals in a stronger sense than they claim. They do acknowledge the existence of the Bohm theory in a footnote, but do not see it as a problem for their non-individualistic view. They write:

\begin{quote}Of course, there is a version of quantum theory, namely Bohm theory, according to which QM is not complete and particles do have definite trajectories at all times. However, Harvey Brown {\em et al.} (1996) argue that the `particles' of Bohm theory are not those of classical mechanics. The dynamics of the theory are such that the properties, like mass, charge, and so on, normally associated with particles are in fact inherent in the quantum field and not in the particles. It seems that the particles only have position. We may be happy that trajectories are enough to individuate particles in Bohm theory, but what will distinguish an `empty' trajectory from an `occupied' one? Since none of the physical properties ascribed to the particle will actually inhere in points of the trajectory, giving content to the claim that there is actually a `particle' there would seem to require some notion of the raw stuff of the particle; in other words {\em haecceities} seem to be needed for the individuality of particles of Bohm theory too. (ETMG, p. 136 fn.)
\end{quote}
Actually, Ladyman and Ross are somewhat one-sided in their reporting of the views of Brown, Elby and Weingard (1996).  For in their paper Brown {\em et al.} are  {\em not} arguing for the view that in the Bohm theory, properties like mass, charge, and so on, normally associated with particles are {\em only} inherent in the $\psi$-field and not in the particles. What they do argue for is that certain experiments (for example, certain types of interferometry experiments) rule out the possibility that these properties are associated with the Bohm particle {\em alone}. 

They point out that there are two principles we can adopt here. Firstly, there is the {\em principle of generosity}, according to which the properties can be attributed to both the $\psi$-field and the particle. Secondly, there is the {\em principle of parsimony} according to which properties such as mass are attributes not of the particle but of the $\psi$-field alone. They do not take a definite stand on which principle we should adopt. However, they draw attention to reasons to adopt the principle of generosity, while at the same time indicating difficulties inherent in the principe of parsimony. It thus seems clear that, {\em contra} what Ladyman and Ross suggest, they are more in favor of the principle of generosity. To be fair, however, we should acknowledge that the issue is subtle and people's views on this vary-- it seems that Harvey Brown himself was in favor of parsimony before opting for generosity. For Brown {\em et al.} (1996) acknowledge that the principle of parsimony is implicit in Brown's earlier work; they also note that it is implied by some of Bell's suggestions (Bell 1990, 30).

Let us now examine in more detail the arguments which suggest that properties such as mass or charge are not inherent only in the particles. In order to bring the issue into focus Brown, Dewdney and Horton (1995) introduce and define the localized particle properties thesis (LPP): particle properties (such as mass, charge etc.) are attributes of the particle rather than the $\psi$-field. That is the mass, say, of the particle is localized at the position of the particle at all times. They go on to point out that several experiments seem to violate the LPP.
 
In the neutron interferometry experiments of Colella, Overhauser and Werner (1975), a neutron stream travels through a beam splitter along two routes, producing an interference pattern.  Now, if the apparatus is tilted in such a way that one of the routes has a higher gravitational potential than the other, the interference pattern is shifted. According to the Bohm theory the particle travels one of the paths while the $\psi$-field travels both paths. Brown {\em et al.} note that if we assume that all of the electron's gravitational mass is concentrated in the path where the particle is, it becomes difficult to understand intuitively why the interference pattern is shifted. For if the empty path $\psi$-field carries no gravitational mass, how could the difference in the gravitational potential integrated over the two paths be felt by the particle? So they argue that for gravitational mass, the LPP seems to be violated.

According to Bohm and Hiley the particle and the  $\psi$-field are strictly speaking indivisible. However we can think of them as two different aspects of an underlying whole. Now, there seems to be no inherent reason in such ontology that, say, the mass should be entirely localized with the particles. Indeed the mathematics suggests that mass is implicated in both the particle and the field aspect.  For mass appears both in the mathematical expression of the kinetic energy of the particle and in the mathematical expression of the quantum potential which reflects the $\psi$-field.  This is in harmony with the principle of generosity, i.e. it seems that mass resides in both the particle aspect and in the field aspect of the individual system.

In the Aharonov-Bohm (AB) effect a similar situation arises with charge. In the AB effect the normal two-slit experimental arrangement has in the geometrical shadow of the two slits a shielded region containing a magnetic field.  The shield is such that the electrons cannot experience the magnetic field directly at all.  Nevertheless one finds that the interference pattern is shifted by an amount that depends on the strength of the magnetic flux in the shielded region.  Since the electrons do not experience this flux directly it is difficult to understand why the interference pattern is changed.  Indeed, Brown, Sj\"{o}qvist and Bacciagaluppi write (1999, p. 234 fn): 
 
 \begin{quote}The expression for the phase shift due to the flux in the shielded solenoid depends on the electronÕs charge being present on spatial loops within the support of the wave function and enclosing the solenoid.
 \end{quote} 
 But again the trajectory of the Bohmian particle associated with the charged particle does not encircle the solenoid. So the LPP seems to be violated for charge.  
 
The AB effect has been numerically analyzed in terms of the Bohm theory by Philippidis, Bohm and Kaye (1982). This analysis shows that in the AB effect,  the vector potential (from which the magnetic line of flux is derived) affects the phase of the wave function in such a way that the latter gives rise to an asymmetric quantum potential. 

\begin{figure}[h]
   \centering
    \includegraphics[width=2in]{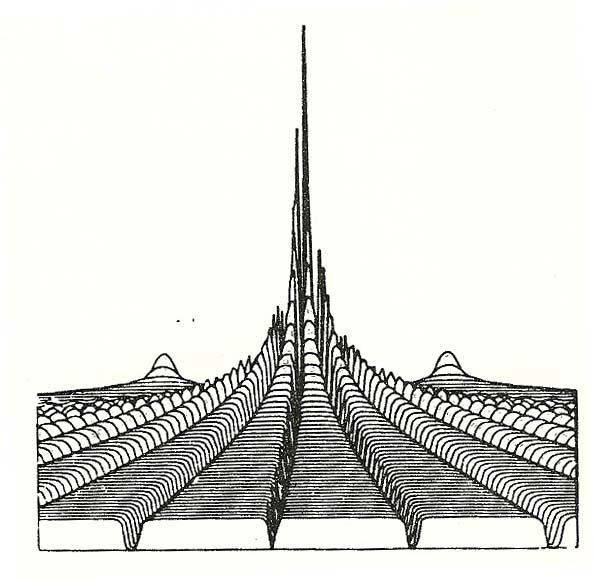}
   \caption{The quantum potential for the Aharonov-Bohm effect.  Notice the asymmetrical shift}
   \label{fig:3}
\end{figure}

If we compare this QPE with that shown in figure 2, where no magnetic line of flux is present, we see that the pattern of the quantum potential has been shifted off the axis of symmetry by an amount that depends on the strength of the enclosed flux. This, in turn, produces a shift in the ensemble of trajectories as was shown in Philippidis, Bohm and Kaye (1982).  This results in an overall pattern shift which has actually been observed in experiment (see Bayh (1962)). As with the gravitational example, it is reasonable to assume that the charge inheres both in the particle and the wave (see Bohm and Hiley 1993: 52-54 for more details).  

\vspace{2cm}

\section{Symplectic symmetry as the common basis of classical and quantum dynamics}

There has been a common misconception that the use of the quantum potential implies a return to  the classical paradigm and that quantum mechanics is a radical departure, so radical that any elements of the classical paradigm must be avoided at all costs.  Indeed the way that the quantum Hamilton-Jacobi equation (2) is derived hides a much deeper relation between classical and quantum dynamics.  Mathematically they have a common kinematic symmetry, namely, the symplectic group of transformations.  As Melvin Brown (2006: v) has succinctly put it,

 \begin{quote} [t]his very general group of transformations maintains the fundamental relationship between position and momentum in mechanics, and its covering group (the metaplectic group) correspondingly transforms the wave function in quantum mechanics.
 \end{quote} 
  
Assuming that the reader may not be familiar with the mathematics of covering groups, we will here approach the subject in a non-formal way (for a more technical presentation, see e.g. de Gosson 2001; Brown 2006). 
  
To bring out this deeper connection, let us return to examine the motivations that led Schr\"{o}dinger to his equation.  When the wave properties of electrons and atoms had been established, Schr\"{o}dinger recalled Hamilton's discussions on the relation between ray optics and wave optics.  Hamilton had shown that paraxial optical rays could be described by the same pair of dynamical equations that he had proposed for classical particles and was attempting to generalise these to capture the wave properties of light. (See Guillemin and Sternberg 1984 for more details.) What the Hamilton-Jacobi theory had shown was that the rays (classical particle trajectories) were perpendicular to a set of surfaces of constant action $S$.  Although the action is well defined mathematically, 
namely, $S=\bm {p.x}-Et$, its physical meaning was not clear. Schr\"{o}dinger noticed that if we could regard these surfaces as surfaces of constant phase, then perhaps, we could find an equation that would form the basis of what he called  ``Hamiltonian undulatory mechanics".

Schr\"{o}dinger's arguments to derive his equation were, at best, heuristic.  Even Schr\"{o}dinger (1926) himself writes ``I realise that this formulation is not quite unambiguous".  But this should not be surprising as the necessary mathematics of the geometry that underlies both classical and quantum dynamics did not exist the 1920s. 

However the equation quickly gave results that agreed with experiment, so that the equation was taken as an {\em a posteriori} given, independent of its origins.  It became {\em the} defining equation of quantum phenomena and as a consequence of trying to understand this equation, the `wave function' became the centre of attention (this attention continues to the present day, see e.g. Albert and Ney ed. 2013).  With this position came the paradoxes of quantum theory that remain unresolved.  Relatively few physicists or philosophers have attempted a sustained exploration of  the deeper mathematical background from which the equation appears.  Indeed the Schr\"{o}dinger equation is taken as a given, arising as if by magic.  Even Feynman (1963) acknowledged that the equation was not derived from anything known in physics or mathematics. As  he remarked: ``It came out of the mind of Schr\"{o}dinger".

The connection between classical and quantum dynamics begins to emerge as we examine the common symmetry, the symplectic symmetry, underlying both Hamiltonian dynamics and the Schr\"{o}dinger dynamics (de Gosson 2001 and 2010). These deeper connections have emerged relatively recently in the mathematics literature and are just beginning to become appreciated by the physics community. What one learns is that classical mechanics and ray optics emerge at the level of the symplectic group of transformations itself, but wave theory and quantum mechanics begin to emerge at a higher level, namely in  the double cover of the group, i.e. the metaplectic group and its generalisation.  The ``double" here means, roughly, that there are always two elements in the metaplectic group representing one element in the symplectic group.

The importance of the double cover is a vital part of quantum mechanics, since it enables one to capture global properties of the type that occur in our discussion of the AB effect. These global properties play a significant role in Bohm's notion of ``unbroken wholeness" which we will discuss in a later section. 

We are already very familiar with this idea of a double cover in the case of rotational symmetry. Here the double cover is the spin group. This gives us the spinor with which we describe fermions and these spinors form the mathematical basis of the entangled singlet state given by equation (1) above. Furthermore it is the relation between the group and its double cover that gives us a mathematical description of the experimentally confirmed difference between a $2\pi$ and a $4\pi$ rotation that shows up in experiments with fermions (see Werner et al. 1975).

This is not the place to go into mathematical details and we will simply point out some interesting results to bring out the unexpected connection between classical and quantum mechanics. Firstly de Gosson and Hiley (2011) have shown that if we formally ``lift" a Hamiltonian flow onto the double cover space, we find a unique flow in this covering space and this flow satisfies a Schr\"{o}dinger-like equation, confirming the results in Guillemin and Sternberg (1984). It is `Schr\"{o}dinger-like' because at this stage it contains an arbitrary parameter with dimensions of action to enable us to put position and momentum on the same footing. The mathematics alone does not enable us to identify this parameter with Planck's constant. Its value is determined by experiment as it is in the standard theory. 

Secondly de Gosson (2010) draws attention to a deep topological theorem in symplectic geometry known as the `Gromov no-squeezing' theorem. This states that even in classical mechanics, it is not possible to reduce a canonical volume such as $\Delta x\Delta p$ by means of a Hamiltonian flow alone. When this region is lifted into the covering space it provides the source of the uncertainty principle. In this way one can say that the symplectic features of the classical world contain the footprint of the uncertainty principle even at the classical level. This structure provides a rigorous mathematical background for Schr\"{o}dinger's `undulatory mechanics'.

\section{Bohmian quantum individuals in the light of the metaplectic group of transformations}

We cannot go into the mathematical details here, nevertheless we feel it is necessary to explain some aspects that emerge from the details.  Since we want to focus on the relationship between the action surfaces and phase surfaces, it is convenient to describe the dynamical evolution in terms of a notion of a flow.  For classical particles, the Hamilton flow is in phase space
\begin{eqnarray*}
z(t')=f_{t',t}z(t)
\end{eqnarray*}
where $z(t)=(x(t),p(t))$ are the coordinates of the particles in the phase space.  The Schr\"{o}dinger flow is 
\begin{eqnarray*}
\psi(t')=F_{t',t}\psi(t)
\end{eqnarray*}
where $\psi(x,t)= (R(x,t), S(x,t))$.  In each case the flow is determined by the Hamiltonian.  By focussing on the flow, we have a different way of understanding the relationship between the two types of behaviour.  

We can illustrate how the flow determines the behaviour of the individual by re-examining the AB effect.  Here the flow must reflect the difference between those paths that encircle the enclosed flux and those that do not. What this means is that the Schr\"{o}dinger flow itself is a global flow, not a local flow. In order to access the relevant global properties of the environment, the flow must carry the information about the global properties of the environment, which is encoded in the covering group.  When we examine the effect of this covering group in the underlying group we find that the energy associated with the particle splits into two parts
\begin{eqnarray*}
E_{\mbox{particle}}=\frac{(\nabla S)^2}{2m}-\frac{\hbar^2}{2m}\frac{\nabla^2R}{R}=KE+QPE.
\end{eqnarray*}
 In this sense the QPE is not the source of a force acting on the particle. Rather it is a potentiality for the behaviour of the particle-like process that finds itself at a particular region in space (this may be somewhat similar to the ideas of Esfeld et al. (2013) who consider the prospects of grounding the motions of particles in dispositions in the Bohm theory). 

The splitting of the energy of the individual is not mere speculation as experiments using weak measurements are in progress to measure these parts (Flack and Hiley 2014).  In view of this radical possibility let us look at the role of the individual in a different way. Recall that Hamilton considered the optic ray as the locus of some invariant independent and indivisible feature of the energy flow, namely the the wave-packet that we now call the photon.  When we apply the same idea to the Schr\"{o}dinger particle, we can regard the `trajectory' as the locus of some indivisible but {\em not necessary localised} energy which we call the `particle'.  In the case of the photon it is not possible to slow it down or give it a precise position so the notion of a `photon trajectory' is questionable.  On the other hand for the quantum particle, it is possible to slow it down and examine it in the classical limit, and show it becomes a point-like object.  Such a situation arises whenever the quantum potential energy (QPE)  is small compared with classical energy. 

Hiley and Mufti (1995) have given a simple model that nicely illustrates this feature.  Suppose we have a situation in which the quantum potential is time dependent and becomes smaller as time progresses. One can then show that the ensemble of quantum trajectories merge smoothly into an ensemble of classical trajectories. Thus a particle following a trajectory in the quantum domain will become a particle obeying the rules expected of a classical particle. In our view these results present strong evidence that it is a coherent possibility that a particle keeps its identity and individuality in a quantum context, even though some of its energy is now involved in exploring its environmental neighbourhood via the $\psi$-field producing the associated QPE.  

The radically different nature of individuality in the quantum domain appears when two or more particles become `entangled' as illustrated in equation (1). In the Bohm approach these particles are coupled by a common quantum potential energy. This potential is {\em non-local} in the sense that the behaviour of one particle is `locked' into the behaviour of the other. This gives a time evolution that involves the two spatially separated particles behaving as a single entity. An example of such an object is the Cooper pair, the electron pair responsible for superconductivity.  It is tempting to see such an entity as new type of individual, where we find a ``twoness" in an underlying individual whole. This is in contrast to two particles described by a product of two wave functions, which can be seen as separate individuals.

If we return to examine the details of the pair of particles described by equation (1), we find an ensemble of correlated trajectories which have been calculated by Dewdney {\em et al.} (1986, 1987, 1988). There we see that if one of the particles enters the field of a Stern-Gerlach magnet, it is then deflected either `up' or `down' depending on the positions of each particle at the time just before the particle enters the magnetic field. The particle in the field has its trajectories changed while the other particle continues in a straight line. At the same time both  spin components become well defined.  This is a surprising result, but shows quite clearly that the individual parts cannot be thought of as isolated `little spinning spheres', a point that was emphasised by Weyl (1931).  

In the AB situation, an ensemble of incident particles would then give rise to a shifted interference pattern, the shift being determined by the enclosed magnetic flux. Nowhere do we lose the identity of the individual particle even though it is responding to a global situation. This notion of individuality is strengthened when it is realised that if the particle were not charged, then its Schr\"{o}dinger evolution would be very different and the interference pattern would be unaffected by the presence of any enclosed magnetic line of flux. In this sense the quantum potential energy is a `private' energy; it is `individual' in the sense of belonging to the individual particle.  This is why on detection the particle reveals its total energy. 
 
Note that this is an example where quantum theory seems to involve stronger and more peculiar individuality than classical physics! This point is often left unnoticed in discussions of individuality in the quantum theory where one typically emphasizes the non-individualistic aspects of the quantum domain. Note in particular that if two particles are conventionally described by a product of two wave functions and the particles do not interact through a classical potential, they do not experience each otherÕs quantum potential even though they may both be in a region of space where their wave functions have significant spatial overlap (see Brown {\em et al.}1996: 313-4).

\section{Problems with the Bohm theory and their solution via the hypothesis of active information}

Bohm was not entirely happy with his 1952 theory, and in his 1957 book  {\em Causality and Chance in Modern Physics}, as well as in a 1962 article (republished in Bohm 1980) he summarized his criticisms.  Firstly, he admitted that the form of the quantum potential is ``strange and arbitrary" and that unlike, say, the electromagnetic field, it has no visible source.  He added that while the theory is logically consistent, the quantum potential should be seen as ``at best a schematic representation of some more plausible physical idea to which we hope to advance later, as we develop the theory further" (Bohm 1980: 102).  He also pointed out that for the many-body system the wave-function lives in a 3N-dimensional configuration space (where N is the number of particles).  It is difficult to understand what such a multi-dimensional $\psi$-field means from a physical point of view (this is a common traditional criticism of the de Broglie-Bohm approach, see e.g. Putnam 1965).  

One way of responding to these problems is an approach to Bohm theory that has become known as ``Bohmian mechanics" which we briefly mentioned above.  Here one can say that the positions of particles are the ``primitive ontology". What, then, is the wave function?  Goldstein and Zanghi (2013) propose that we consider the wave function as nomological, something more in the nature of a law than a concrete physical reality. So if the wave function does not describe a physical field, the question of the 3N-dimensional field for the many-body system does not arise.  However, we believe that quantum potential approach enables us to explore the ontological meaning of quantum theory in a deeper way, so let us consider whether it is possible to make this approach physically more viable. 

Remember the two worries: the form of the quantum potential is strange and arbitrary, and it is difficult to give a physical interpretation to a multi-dimensional $\psi$-field.  When Bohm re-examined his 1952 theory in the late 1970s he realized that the quantum potential might be telling us something radically new about the nature of reality.  For he noticed that the quantum potential depends only upon the form, or the second spatial derivative of the amplitude R of the $\psi$-field.  This form, in turn, reflects the form of the environment (such as the presence of slits, but also features such as magnetic flux lines and differences in gravitational potential).  

Could it be that the $\psi$-field is literally ``in-forming" or putting form into the activity of the particle, rather than pushing and pulling the latter mechanically?   Bohm called such information ``active information", because this is an instance where information acts to bring about changes in the behaviour of the particle. Note also that this idea of the $\psi$-field as a field that encodes information provides a new way of understanding the multi-dimensionality of the $\psi$-field for the many-body system.  Indeed, Bohm suggested that  the wave function describes not a multidimensional field, but rather an information structure that can quite naturally be considered to be multidimensional, i.e., organized into as many sets of dimensions as may be needed. Thus the quantum wave field is not regarded as a simple source of a mechanical force. He speculated that the information encoded in the quantum potential is carried in some much more subtle level of matter and energy which has not yet manifested in physical research (see Bohm and Hiley 1987: 336). Note that the two key anomalies of the 1952 theory (i.e., the arbitrary form of {\em Q} and the multidimensionality of the many-body $\psi$-field) thus became the corner stones of a new interpretation of the $\psi$-field as a field of information (Pylkk\"{a}nen 1993).

Let us next consider the individuality of Bohmian particles in the light of this active information approach.  First of all, Bohm and Hiley were led to propose that, say, an electron has an internal structure which enables it to respond to the information in the $\psi$-field.  So, according to this view Bohmian quantum particles are not the point particles of classical physics, but much more subtle entities.  Note also that because of the holistic features of the quantum potential, these particles have only a relative autonomy. In the case of a single particle, because the quantum potential only depends upon the form of the $\psi$-field, it does not necessarily fall off with distance even if the intensity of the $\psi$-field becomes weak as the field spreads out.  This means that even very distant features of the environment (e.g. slits) can have a strong effect upon the particle, thus underlining the lack of its autonomy.  Strictly speaking the entire experiment has to be treated as an undivided whole, which is reminiscent of Bohr's view.  However, while Bohr suggested that this whole is unanalyzable, in the Bohm theory one can now analyze it in thought in terms of the movement of the particle acted on by the quantum potential.

In the two-body system the autonomy of the individual becomes weaker still, for the quantum potential depends on the position of both particles in a way that does not necessarily fall off with the distance. This means that there is the possibility of a non-local interaction between the two particles. We can generalize this to the N-body system where the behaviour of each particle may depend non-locally on all the others, no matter how far away they may be.  Non-locality is an important new feature of the quantum theory, but Bohm used to emphasize that there is yet another feature that is even more radical. For in the Bohm theory there can be a non-local connection between particles that depends on the quantum state of the whole, in a way that cannot be expressed in terms of the relationships of the particles alone.   This quantum state of the whole, described by the many-body wave function, evolves in time according to the Schr\"{o}dinger equation, which led Bohm and Hiley to write:

\begin{quote}
Something with this sort of independent dynamical significance that refers to the whole system and that is not reducible to a property of the parts and their inter-relationships is thus playing a key role in the theory. ... {\em this is the most fundamental new ontological feature implied by quantum theory}. (Bohm and Hiley 1987: 332)
\end{quote}

The above quote reveals the holistic character of Bohm's interpretation of quantum theory.  Even if his 1952 theory in a sense rediscovered the lost individuality of quantum objects, his quantum ontology was not a return to the individuals of classical physics.  He thought that quantum theory was primarily about dynamical wholeness that is not reducible to the interactions between individuals.  As Max Jammer has pointed out, this means that the individuals are not ``constitutive" to the whole but rather depend on the state of the whole (1988: 696). Related to this, Tim Maudlin has commented:

\begin{quote}
David Bohm has long contended that what is radically new about the quantum theory is the ``undivided wholeness" that it posits, and if Bohm is right, philosophical commentaries on the quantum theory have long been preoccupied with the wrong features of the theory. (1998: 49)
\end{quote}

What is also relevant here for the present volume is that the wholeness of Bohmian quantum systems seems analogous to the organic unity of biological systems, suggesting interesting links between physics and biology:

\begin{quote}
...the quantum potential arising under certain conditions has the novel quality of being able to organize the activity of an entire set of particles in a way that depends directly on the state of the whole. Evidently, such an organization can be carried to higher and higher levels and eventually may become relevant to living beings. (Bohm and Hiley 1987: 332)
\end{quote}

The quantum potential approach thus provides potentially useful tools for a holistic approach in biology.  There seems to be at least an analogy in the way the whole and the part are related in some quantum phenomena and some biological phenomena.

Our discussion above suggests a richer view of Bohmian particles than is presupposed by Ladyman and Ross, or Brown {\em et al}. Indeed, Bohm proposed that it is plausible that the behaviour and structure of matter does not always become simpler as we go to lower dimensions. Radically, he suggested that a particle such as an electron may have a structure (somewhere between 10-16 cm and 10-33 cm).  This structure is assumed to be complex and subtle enough to respond to the information described by the wave function.  So according to this hypothesis, there is definitely more to the individuality of Bohmian particles than mere {\em haecceities}.

\section{Bohm's scientific structuralism}

We have seen above that while the 1952 Bohm theory in a sense rediscovered the quantum individuals and particle trajectories that got lost in the usual interpretation, the theory should by no means be seen as a return to a mechanistic ontology where individuals are fundamental and constitutive to the whole.  Bohm's other main line of research, known as the ``implicate order", likewise sees quantum theory as a guide towards a conception of a new holistic and dynamic order in physics (Jammer 1988: 696). This work which begins to develop in the early 1960s, aims to develop a deeper underlying theory from which quantum theory and relativity can be derived as approximations, and their relation thus understood.  This framework suggests a strongly structuralist, process-oriented way of understanding individual quantum systems which is in some ways similar to Ladyman and Ross's structural realism. At an early phase of this work Bohm wrote: 
 
 \begin{quote}
 In this theory ... the notion of a separately existing entity simply does not arise. Each entity is conceptually abstracted from a totality of process... with the electron, what actually exists is a structure of underlying elementary processes or linkages supporting a pattern corresponding to an electron. (1965a: 291).
 \end{quote}
  In the later implicate order view, an electron is not a little billiard ball that persists and moves, but should more fundamentally be understood as 
  
  \begin{quote}...a recurrent stable order of unfoldment in which a certain form undergoing regular changes manifests again and again, but so rapidly, that it appears to be in continuous existence (Bohm 1980: 194). 
 \end{quote}
  
  Finally, in the final chapter of their 1993 book {\em Undivided Universe} Bohm and Hiley, when discussing quantum field theory and emphasizing the ontological primacy of movement required by relativity,  summarize this non-individualistic line of thought as follows: 
  \begin{quote}
  ...the essential qualities of fields exist only in their movement [...] The notion of a permanently extant entity with a given identity, whether this be a particle or anything else, is ... at best an approximation holding only in suitable limiting cases. (1993: 357). 
 \end{quote} 
  Thus much of Bohm's work supports the idea that individuals are not metaphysically fundamental in the light of contemporary physics.  His emphasis on notions such as ``structural process" (1965b), ``order" and ``movement" (1980) as fundamental in physics suggests that the philosophical home of Bohm (and Hiley's) more general approach to physics might well be found in some form of scientific structuralism which takes movement as fundamental, rather than in a metaphysics which takes individuals as basic (cf. Ladyman and Ross 2007).   Indeed, Hiley's recent work on symplectic geometry can be seen as bringing Bohm's 1952 approach closer to scientific structuralism.  For ultimately Hiley' s work leads to the algebraic approach that was initiated by Bohm and Hiley (1993, ch 15).

\section{Concluding remarks.}

We have above shown that the prospects of individuality in the Bohm theory are stronger than Ladyman and Ross imply.  This  suggests that there is an underdetermination of metaphysics by physics in non-relativistic quantum theory when it comes to the question of individuality (as indeed has been emphasized by French and Krause 2006: 189-197). However, it is important to realize that the notion of an individual in the Bohm theory  - especially in Bohm and Hiley's (1987, 1993) developed account of it - is very different from what we would expect from the classical perspective.  For although the Bohmian quantum individual has a well defined energy, that energy is not a local energy.  This is consistent with Niels Bohr's views in two ways. Firstly, as emphasized in Bohm and Hiley (1993),  the particle is never separated from the quantum field.  It is an invariant feature of the total underlying process. This is consistent with Bohr's notion of the ``impossibility of subdividing  quantum phenomena" in the sense that the whole experimental arrangement must be taken into account (Bohr 1958:50-51).
 
 Secondly, we have suggested that the Bohmian individual is not a localised point-like object.  As Bohr remarks (1958:73) the quantum process is a ``closed indivisible phenomenon".  The energy is not localised at a point.  In fact complementarity can be taken to imply that energy transcends space-time.  Nevertheless there is a centre of energy, a generalisation of the centre of mass which can be given a position in space-time.  It is this particle-like centre that moves with the Bohm momentum.

These ideas are not consistent with a classical notion of a particle and we feel can only be given a more comprehensive meaning in terms of something like Bohm's (1965b) notion of ``structural process".  Thus the overall Bohmian approach to physics does not, from the metaphysical point of view, mean a return to the individuals of classical physics, but has instead strongly structuralist features. In particular, we noted that Bohm and Hiley have since the 1960s been developing a broader scheme they call ``the implicate order", which goes beyond the 1952 Bohm theory (Bohm 1980; Bohm and Hiley 1993: ch15; Hiley 2011; Pylkk\"{a}nen 2007; for Bohm's own attempt to reconcile "hidden variables" and the implicate order, see his 1987). We acknowledge that this scheme seems to have some relevant similarities to Ladyman and Ross's ontic structural realism, while there also may be some significant differences. The discussion of these similarities and differences will, however, be a subject of another study (some preliminary attempts have already been made by P\"{a}ttiniemi 2011 and Pylkk\"{a}nen 2012).

\section{Acknowledgements}

We received extensive and insightful comments on this paper from Guido Bacciagaluppi and David Glick for which we are grateful.   Bacciagaluppi's comments encouraged us to strengthen our characterization of the particle pair in an entangled state as an individual, while Glick's comments prompted us to emphasize the differences between Bohmian and classical particles.  We also received valuable feedback from Alexander Guay, Tuomas Tahko and an anonomous referee. An early version of this paper was presented May 14th 2012 at the Philosophy of Science seminar of the University of Helsinki and May 19th 2012 at the Individuals Across Sciences -symposium in Paris.  We thank the participants of these events for the many thoughtful comments we received.



\bibliography{myfile}{}
\bibliographystyle{plain}

\end{document}